\documentclass[aps,prb,reprint,superscriptaddress]{revtex4-2}
\usepackage{amsfonts}
\usepackage{amssymb}
\usepackage{amsmath}
\usepackage{graphicx}
\usepackage{epsfig}
\usepackage{array}
\usepackage{braket}
\usepackage{multirow}
\usepackage[table,xcdraw]{xcolor}
\usepackage[colorlinks]{hyperref}
\usepackage{color}
\usepackage{hyperref}

\hypersetup{
	colorlinks=true,
	linkcolor=red,
	filecolor=blue,
	urlcolor=blue,
	citecolor=blue,
}

\begin{document}

\title{Origin of Moir\'{e} Potentials in WS$_2$/WSe$_2$ Heterobilayers: Contributions from Lattice Reconstruction and Interlayer Charge Transfer}
\author{Youwen Wang}
\author{Nanya Gao}
\author{Qingjun Tong}
\email{tongqj@hnu.edu.cn}
\affiliation{School of Physics and Electronics, Hunan University, Changsha
	410082, China}

\begin{abstract}
Moir\'{e} superlattices formed in WS$_2$/WSe$_2$ heterobilayers have emerged as an exciting platform to explore the quantum many-body physics. The key mechanism is the introduction of moir\'{e} potentials for the band-edge carriers induced by the lateral modulation of interlayer interactions. This trapping potential results in the formation of flat bands, which enhances the strong correlation effect. However, a full understanding of the origin of this intriguing potential remains elusive. In this paper, we present a comprehensive investigation of the origin of moir\'{e} potentials in both R-type and H-type moir\'{e} patterns formed in WS$_2$/WSe$_2$ heterobilayers. We show that both lattice reconstruction and interlayer charge transfer contribute significantly to the formation of moir\'{e} potentials. In particular, the lattice reconstruction induces a nonuniform local strain, which creates an energy modulation of ${\sim}200\ \mathrm{meV}$ for the conduction band-edge state located at WS$_2$ layer and ${\sim}20\ \mathrm{meV}$ for the valence band-edge state located at WSe$_2$ layer. In addition, the lattice reconstruction also introduces a piezopotential energy, whose amplitude ranges from ${\sim}40$ to ${\sim}90\ \mathrm{meV}$ depending on the stacking and band-edge carrier. The interlayer charge transfer induces a built-in electric field, resulting in an energy modulation of ${\sim}80\ \mathrm{meV}$ for an R-type moir\'{e} and ${\sim}40\ \mathrm{meV}$ for an H-type moir\'{e}. Taking into account both effects from lattice reconstruction and interlayer charge transfer, the formation of moir\'{e} potential is well understood for both R-type and H-type moir\'{e}s. This trapping potential localizes the wavefunctions of conduction and valence bands around the same moir\'{e} site for an R-type moir\'{e}, while around different moir\'{e} site for an H-type one. Our work not only provides an efficient method to study lattice-mismatch induced moiré patterns, but also gives help in understanding the intriguing moir\'{e} physics in WS$_2$/WSe$_2$ heterobilayers.

\end{abstract}

\maketitle

\section{Introduction}

van der Waals (vdW) heterostructures provide a powerful approach to engineer novel physics and functional devices \cite{Geim2013,Liu2016,Kennes2021}. A characteristic feature of vdW heterostructures is the presence of a moiré pattern because of the lattice mismatch and/or twisting \cite{He2021}. This moir\'{e} superlattice formed in magic-angle twisted bilayer graphene has led to the experimental discovery of superconducting and correlated insulating states \cite{Cao20181,Cao20182}. Another frontier of recent studies is the  moir\'{e} superlattice formed in transition metal dichalcogenides (TMDs), which have many advantageous optoelectronic properties \cite{Wilson2021,Huang2022,Mak2022,Regan2022}. Because of the strong Coulomb interaction, moir\'{e} patterns formed in WS$_2$/WSe$_2$ heterobilayers are of particular interest, which have emerged as a rich playground for exploration of novel excitonic and electronic states, such as moir\'{e} excitons \cite{Jin2019,Jin20192,Sunx2022,Gec2023,Huang2025}, Mott and generalized Wigner crystal states \cite{tang2020,Regan2020,shimazaki2020,nuckolls2024,xu2020,huang2021,jin2021,liu2021,miao2021,chen2022,zhang2022,gu2022,wang2023}. The lattice mismatch between WS$_2$ and WSe$_2$ is about $4\%$, which gives a moir\'{e} pattern of ${\sim}8\ \mathrm{nm}$ \cite{Liu2015,Li2021,zhao2024}. The electronic structure of a WS$_2$/WSe$_2$ heterobilayer is of type-II band alignment, where the conduction and valence bands are located at WS$_2$ and WSe$_2$ layer respectively, and the excitonic ground state is of an interlayer configuration \cite{Kang2013,Rivera2015,Jiang2021}. Because of the breaking of inversion symmetry in the monolayer TMDs, the formed moir\'{e} pattern has two types, named as R-type and H-type respectively \cite{Tong2017}. Furthermore, recent experimental studies have shown that both in-plane and out-of-plane lattice reconstructions in WS$_2$/WSe$_2$ heterobilayers are remarkable, which are believed to impact greatly their optical and electronic properties \cite{Li2021,zhao2024}.

The key to the emergence of novel physics in moir\'{e} superlattices lies in the formation of moir\'{e} potential, which results in nearly flat minibands with localized electronic and excitonic states \cite{Wu20182,Yu2017}. In a long-period moiré pattern, the interlayer atomic configuration locally resembles lattice commensurate structures while changes smoothly over long range \cite{Tong2017,Jung2014}. Because the interlayer interaction generally depends on the stacking order, such a spatial change of interlayer configurations then introduces a spatial modulation of interlayer interaction in the moiré pattern \cite{park2015,xia2015,yu2015,zhang2017,sivadas2018,jiang2019}. This lateral modulation of interlayer interactions then defines a trapping potential (dubbed moir\'{e} potential) for the band-edge electronic and excitonic states. It is believed that there are multiple factors that would contribute to the moir\'{e} potential. In particular, interlayer hybridization would couple the two layers, which results in a renormalization of their band structures \cite{Yu2017,Wang2017}. Furthermore, the lattice reconstruction within each layer induces a local strain, which would change the band structure of each monolayer \cite{Shabani2021,Yang2024}. In twisted homobilayers or chalcogen-matched heterobilayers, it has been shown that the lattice reconstruction would also introduce a piezoelectric effect that strongly confines band-edge carriers \cite{Enaldiev2020,Enaldiev2D2021,Magorrian2021,Ferreira2021}. Finally, the interlayer charge transfer induces a built-in electric field, which would shift relatively the band structures of the two layers \cite{Tong2021}. However, a comprehensive understanding that takes into account all of these effects and accounts for different moir\'{e} types is still lacking.

\begin{figure*}[tbp]
	\begin{center}
		\includegraphics[width=0.8\textwidth]{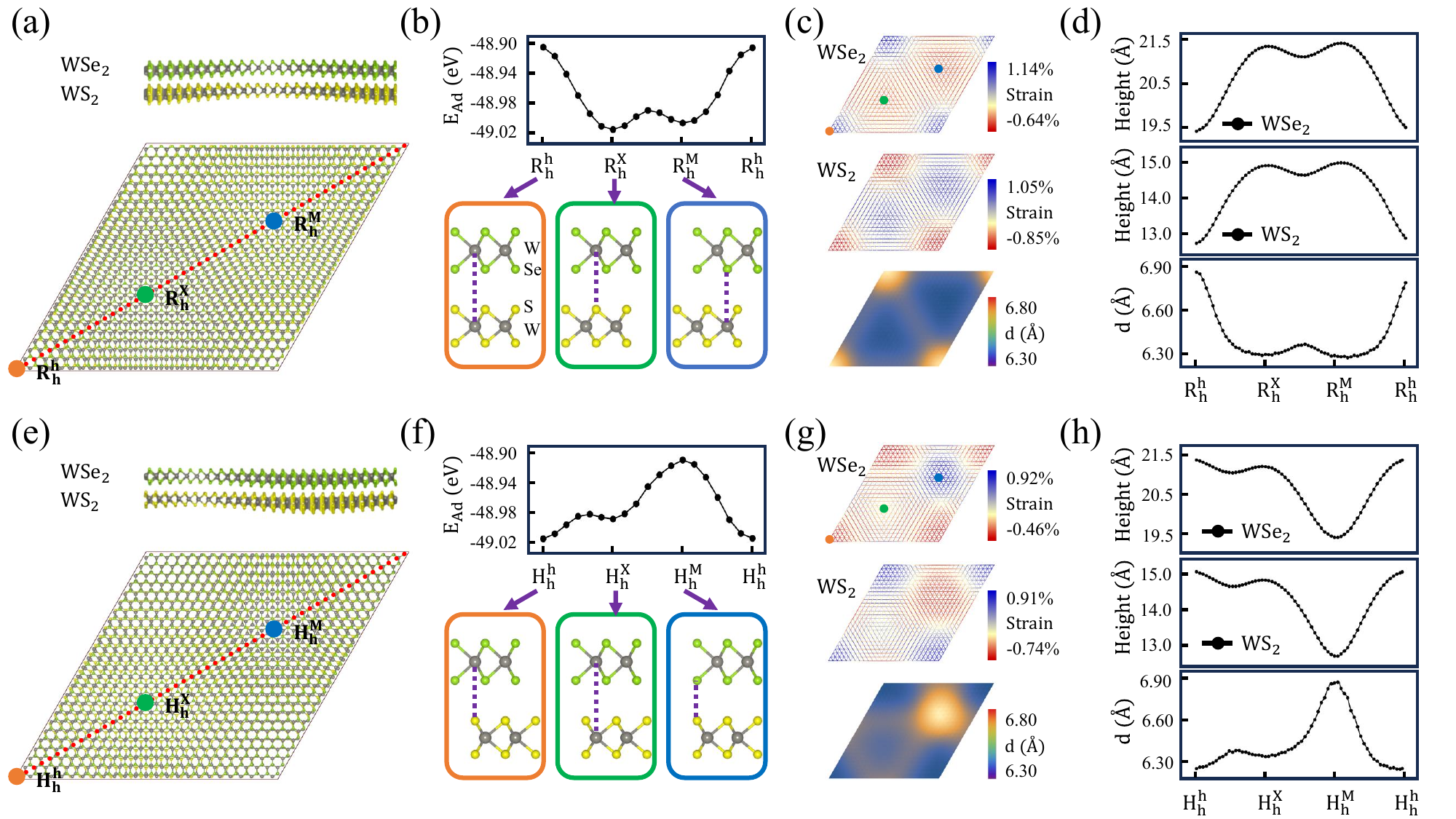}
	\end{center}
	\caption{(a) Relaxed atomic structure of an R-type moir\'{e} pattern formed in WS$_2$/WSe$_2$ heterobilayers. The top panel is a lateral view showing strong out-of-plane corrugation. The three high-symmetry locales along the long-diagonal of a moir\'{e} unit cell are labeled as $R^h_h$, $R^X_h$, and $R^M_h$. The moir\'{e} periodicity is 8.25 nm formed by a lattice mismatch of $4\%$ between WS$_2$ and WSe$_2$ with no relative rotation. (b) Adhesion energy as a function of interlayer stacking along the long diagonal of the relaxed moir\'{e} pattern in (a). The interlayer atomic configurations at three high-symmetry locales are given in the lower panel. (c) The extracted lattice-reconstruction induced in-plane strain distribution in the WSe$_2$ layer (upper) and WS$_2$ layer (middle). The interlayer distance $d$ is given in the lower panel, which is defined by the distance between the two W atomic planes. (d) Out-of-plane corrugation of WSe$_2$ layer (upper) and WS$_2$ layer (middle) and the interlayer distance (lower) along the long diagonal of the relaxed moir\'{e} pattern. (e)-(h) The same as (a)-(d) for the an H-type moir\'{e} pattern, where the upper WSe$_2$ layer has a rotation of 180 degree relative to the lower WS$_2$ layer.}
	\label{fig1}
\end{figure*}

In this work, we present a comprehensive study on the origin of moir\'{e} potentials in both R-type and H-type moir\'{e} patterns formed in WS$_2$/WSe$_2$ heterobilayers. We find that both types of moir\'{e} show strong lattice reconstruction in both in-plane and out-of-plane directions. From the calculation of the miniband structures, we show that in the R-type moir\'{e}, the conduction and valence band-edge states are trapped around the same moir\'{e} site. While in the H-type moir\'{e}, they are trapped around different moir\'{e} sites. This different localization behavior dates back to the distinct morphology of moir\'{e} potentials of the two types of moir\'{e}, which is contributed by lattice-reconstruction induced local strain and interlayer charge transfer. In particular, we find that, for both R-type and H-type moir\'{e}s, the lattice-reconstruction induced local strain results in a strong energy modulation with an amplitude of ${\sim}200\ \mathrm{meV}$ for the conduction band-edge states, while it is of nearly an order smaller for the valence band-edge states. In addition, the lattice reconstruction also introduces a piezopotential energy, whose amplitude ranges from ${\sim}40$ to ${\sim}90\ \mathrm{meV}$ depending on the stacking and band-edge carrier. We also find significant interlayer charge transfer effect in both types of moir\'{e}, which leads to band modulations with an amplitude of ${\sim}80\ \mathrm{meV}$ for the R-type moir\'{e} and ${\sim}40\ \mathrm{meV}$ for the H-type moir\'{e}. Taking into account the effects from local-strain induced band modulation, piezopotential energy and interlayer charge transfer, the localization behavior of band-edge states in both R-type and H-type moir\'{e}s can be properly explained.  

The rest of the paper is organized as follows. In Sec. \ref{atomic}, we give the atomic structures of the moiré patterns in both R-type and H-type WS$_2$/WSe$_2$ heterobilayers and illustrate their detailed lattice reconstruction effect in both in-plane and out-of-plane directions. The miniband structures of the moir\'{e} patterns are given in Sec. \ref{miniband}. The wavefunctions of the conduction and valence bands and their localization behaviors are also presented. In Sec. \ref{origin}, we discuss the origin of moir\'{e} potentials for both types of moir\'{e} patterns, and explain the localization behavior of band-edge states. Finally, discussions on the relevance of our moir\'{e} potential to recent experiments and a summary are given in Sec. \ref{conclusions}. The details of first-principles calculations are given in the Appendix.

\section{Atomic structures of moir\'{e} patterns in WS$_2$/WSe$_2$ heterobilayers}\label{atomic}
In their monolayer forms, TMDs break the inversion symmetry and possess threefold rotation symmetry with the rotation center located at a hexagon center (h), chalcogen (X) site, and metal (M) site respectively \cite{Liu2015}. As a result, when stacking together, TMD heterobilayers may have two types of moir\'{e} patterns, i.e. with lattice mismatch and/or rotation angle near $0^{\circ}$ (R-type) or $60^{\circ}$ (H-type), as shown in Figs. \ref{fig1}(a) and \ref{fig1}(e). In a long-period moir\'{e}, the interlayer atomic configuration changes smoothly, which can be characterized by a vector $\textbf{r}$ that defined in a unit cell of the monolayer \cite{Tong2017}. In our study of WS$_2$/WSe$_2$ heterobilayers, there exist three moir\'{e} sites with threefold rotation symmetry in both types of moir\'{e}s. We name these moir\'{e} sites as $R^\mu_h$ and $H^\mu_h$ in R-type and H-type stacking respectively, where the $\mu$ site of the WSe$_2$ layer vertically aligned with the $h$ site of the WS$_2$ layer.

Because of distinct interlayer stacking registry, the locally commensurate heterobilayers in the moir\'{e} have different adhesion energies. First-principles calculations given in Figs. \ref{fig1}(b) and \ref{fig1}(f) show that, for R-type WS$_2$/WSe$_2$ moir\'{e}, the $R^X_h$ stacking is the lowest energy configuration (with largest adhesion energy), while $R^h_h$ is the highest one. For H-type moir\'{e}, the $H^h_h$ stacking is the lowest energy configuration, while $H^M_h$ is the highest one. Because of this stacking dependent adhesion energy, the moir\'{e} lattice experiences strong structure reconstruction to reduce total energy. The final relaxed structure is then determined by the competition between the intralayer elastic energy caused by lattice reconstruction and the interlayer adhesion energy \cite{Enaldiev2020,Enaldiev2D2021,Magorrian2021,Ferreira2021,Carr2018,Zhuziyan2020}. We use LAMMPS to perform the structural relaxation with the calculation details given in Appendix. For R-type moir\'{e}, TMD lattice would relax itself to enlarge the $R^X_h$ stacking area and reduce the $R^h_h$ stacking area [Fig. \ref{fig1}(a)]. For H-type moir\'{e}, TMD lattice would relax itself to enlarge the $H^h_h$ stacking area and reduce the $H^M_h$ stacking area [Fig. \ref{fig1}(e)]. We note that, because of this lattice reconstruction, the mapping between the local registry $\textbf{r}$ and moir\'{e} location $\textbf{R}$ becomes nonlinear.

As a result of this structure reconstruction, the moir\'{e} lattices feature both in-plane and out-of-plane deformations in each constituent monolayer. Figures \ref{fig1}(c) and \ref{fig1}(g) show the extracted lattice-reconstruction-induced local strain in each layer. The magnitude of this local strain can reach ${\sim}1\%$ for both R-type and H-type moir\'{e}s. In particular, for R-type case, the local strain is compressive at $R^X_h$ and $R^M_h$ stackings and tensile at $R^h_h$ stacking on the WSe$_2$ layer. While on the WS$_2$ layer, the local strain is opposite. For H-type case, the local strain is compressive at $H^h_h$ stacking and tensile at $H^M_h$ site on the WSe$_2$ layer. While on the WS$_2$ layer, the local strain is opposite. In addition, there exists a strong out-of-plane lattice reconstruction as shown in Figs. \ref{fig1}(d) and \ref{fig1}(h). In particular, for both types of moir\'{e}s, there is a modulation of interlayer distance. The largest distance is at $R^h_h$ stacking for R-type and $H^M_h$ stacking for H-type, both of which have chalcogen atoms sitting directly to each other. More interestingly, we find a strong out-of-plane modulation in each monolayer, both of which in the two types of moir\'{e}s can reach as high as $2$ \AA, which is almost one third of the interlayer distance.

\section{Flat miniband and the localization of band-edge states}\label{miniband}
We now turn to studying the electronic structures of the moir\'{e} patterns formed in WS$_2$/WSe$_2$ heterobilayers, with the focus on the flat bands formed near the band edges and their localization behavior. The calculation details are given in Appendix. Figure \ref{fig2}(a) shows the first-principles calculated miniband structure of an R-type moir\'{e} pattern, with the orbital projections indicated explicitly. A series of nearly flat bands are observed in both conduction and valence bands. Compared with the bandwidths in valence bands, the ones in conduction bands are much smaller, suggesting a more compact trapping effect therein. From the analysis of the orbital components, we find that the conduction band-edge states are mainly composed of $d_{z^2}$ orbital, and the valence band-edge states are mainly composed of $d_{x^2-y^2}$ and $d_{xy}$ orbitals. This suggests that the conduction miniband originates from the conduction band at K point of WS$_2$ layer and the valence miniband originates from the valence band at K point of WSe$_2$ layer \cite{Liu2015}. The distribution of the wavefunctions at the band edges in the moir\'{e} unit cells is given in Figs. \ref{fig2}(b) and \ref{fig2}(c). From the lateral view, we find that the conduction band-edge state is localized at the WS$_2$ layer, while the valence band-edge state is localized at the WSe$_2$ layer confirming the type-II band alignment. The top views show that both the conduction and valence band-edge states are localized around $R^M_h$ site. The former is more localized than the latter, because its miniband is flatter.

\begin{figure}[tbp]
	\centering
	\includegraphics[width = 1.0\columnwidth] {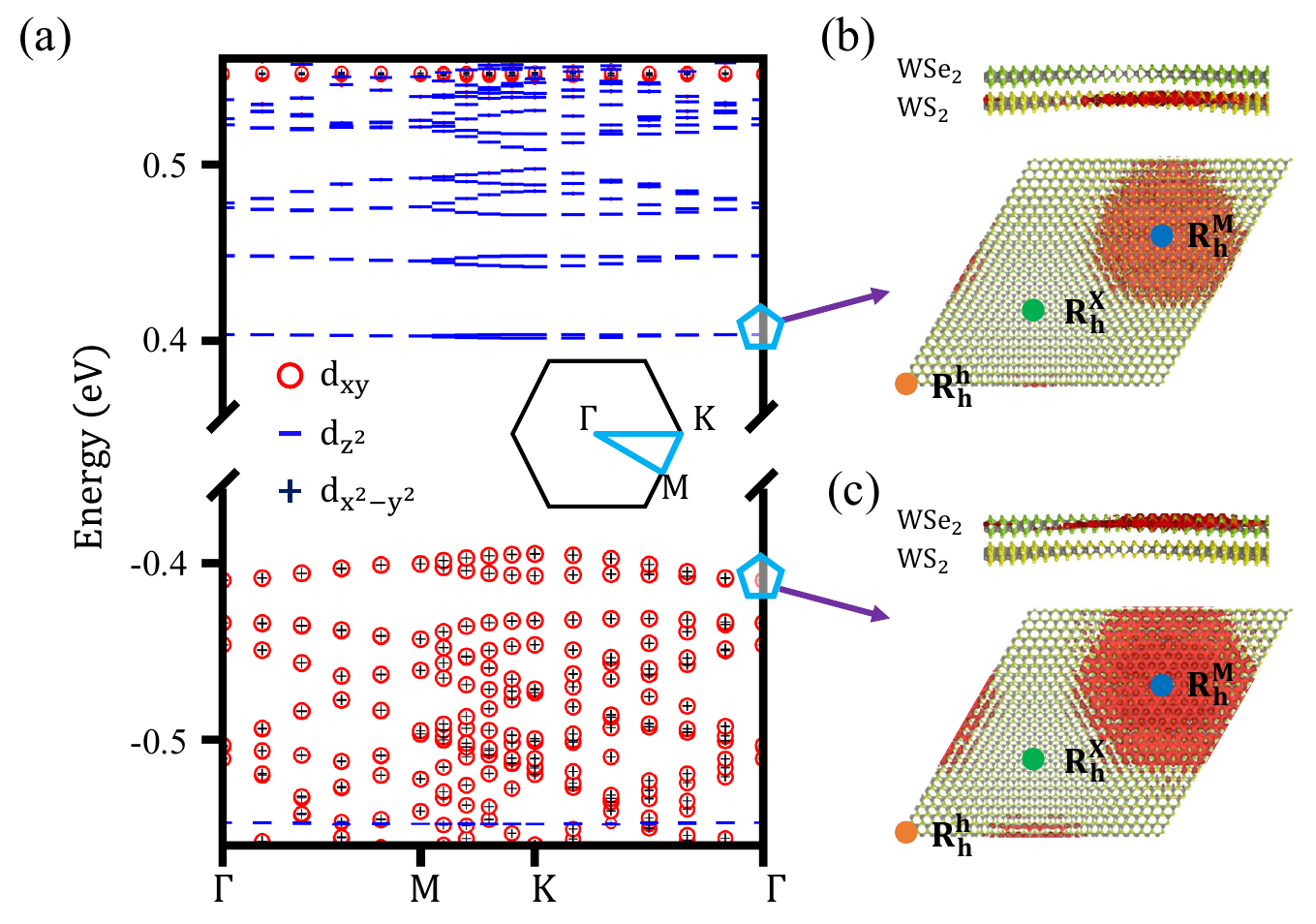}
	\caption{(a) Electronic band structure of a relaxed R-type WS$_2$/WSe$_2$ moir\'{e} pattern along the high-symmetry lines in the hexagonal mini-Brillouin zone (inset). The projected orbital contributions are indicated. The distribution of the wavefunctions in the moir\'{e} unit cell for the conduction (b) and valence (c) band-edge states at the energy points marked by the pentagons in (a), both of which are localized around $R^M_h$ site. The top panels show the lateral views.} \label{fig2}
\end{figure}

\begin{figure}[tbp]
	\centering
	\includegraphics[width = 1.0\columnwidth] {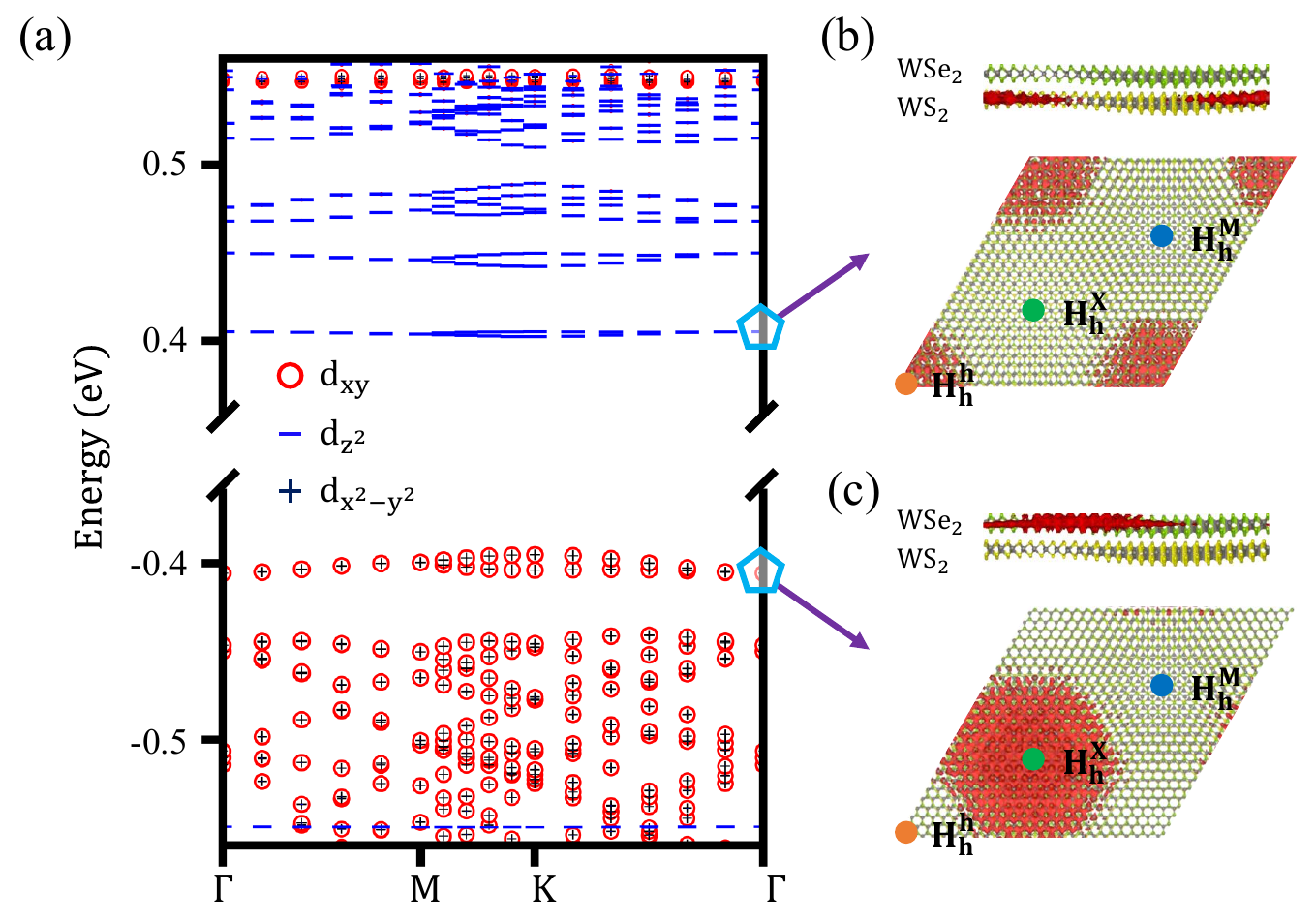}
	\caption{(a) Electronic miniband structure of a relaxed H-type WS$_2$/WSe$_2$ moir\'{e} pattern along the high-symmetry lines in the hexagonal mini-Brillouin zone. The projected orbital contributions are indicated. The distribution of the wavefunctions in the moir\'{e} unit cell for the conduction (b) and valence (c) band-edge states at the energy points marked by the pentagons in (a), which are localized around $H^h_h$ site and $H^X_h$ site respectively. The top panels show the lateral views.} \label{fig3}
\end{figure}

Figure \ref{fig3}(a) shows the band structure of an H-type moir\'{e} pattern, which features flat bands in both conduction and valence minibands similar to the R-type case. The orbital projection suggests that these flat bands also originate from the band-edge states from K point of the WS$_2$ and WSe$_2$ layer respectively. However, different from the R-type moir\'{e}, the wavefunctions of the conduction and valence band-edge states are located around different moir\'{e} sites. In particular, the conduction band-edge state is localized around  $H^h_h$ site [Fig. \ref{fig3}(b)], while the valence band-edge one is localized around  $H^X_h$ site [Fig. \ref{fig3}(c)]. This vertical misalignment of electron and hole wavefunctions is responsible for the experimental observation of intercell moiré exciton complexes with large in-plane electrical quadrupole moments \cite{wang2023}.

\section{Origin of moir\'{e} potentials}\label{origin}
\begin{figure}[tbp]
	\centering
	\includegraphics[width = 1.03\columnwidth] {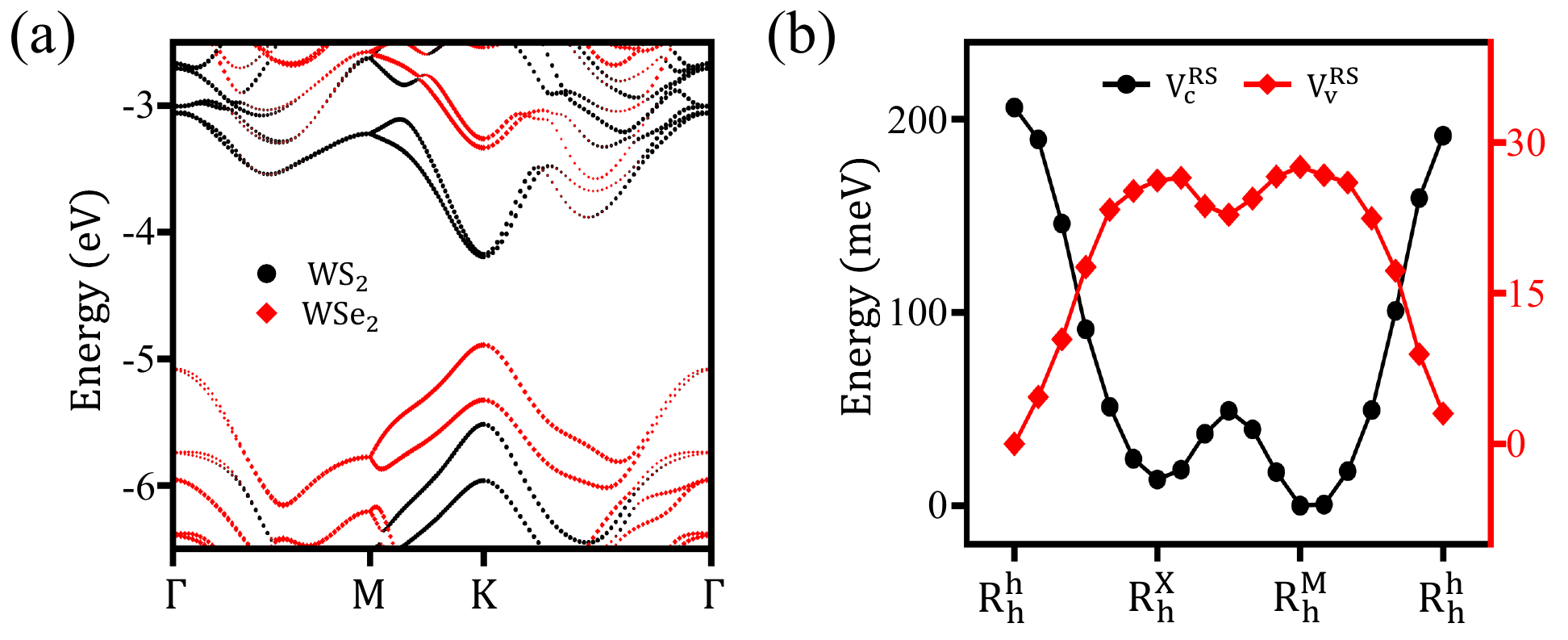}
	\caption{(a) Layer-projected band structure of a WS$_2$/WSe$_2$ heterobilayer. The type-II band edges are located at K point. A commensurate $R^X_h$ stacking is adopted. (b) Modulation of conduction band-edge energy of monolayer WS$_2$ ($V^{RS}_c$ as black line with dots) and valence band-edge energy of monolayer WSe$_2$ ($V^{RS}_v$ as red line with diamonds) as a function of local strain adopted along the long diagonal of Fig. \ref{fig1}(c).} \label{fig4}
\end{figure}

The distinct localization behavior of band-edge states in the two types of moir\'{e} patterns suggests that they have different moir\'{e} potential profiles contributed by various factors. Because of the large band offset (${\sim}860\ \mathrm{meV}$ for the conduction band and ${\sim}630\ \mathrm{meV}$ for the valence band), the interlayer hybridization is significantly quenched in the WS$_2$/WSe$_2$ heterobilayer [cf. the layer-projected band structure in Fig. \ref{fig4}(a)]. We then focus on studying the effects from lattice reconstruction and interlayer charge transfer in the following. We note that the lattice reconstruction mainly affects the constituent monolayers, so it is sufficient to study the effect of lattice-reconstruction-induced local strain by using TMD monolayers. In particular, based on the layer projection shown in Fig. \ref{fig4}(a), we find that it is sufficient to study the local strain effect on the conduction band of monolayer WS$_2$ and valence band of monolayer WSe$_2$ at K point. This adoption of TMD monolayers can avoid the introduction of additional strain (${\sim}\pm2\mathrm{\%}$) to construct commensurate bilayers and also distinguish the contribution of local strain from interlayer charge transfer that happens in a bilayer. In the following, we discuss the R-type and H-type moir\'{e}s separately.

\subsection{R-type moir\'{e}}
Figure \ref{fig4}(b) shows the energy modulations of conduction band of monolayer WS$_2$ ($V^{RS}_c$ as black line with dots) and valence band of monolayer WSe$_2$ ($V^{RS}_v$ as red line with diamonds) as a function of local strain extracted from the relaxed R-type moir\'{e} pattern as shown in Fig. \ref{fig1}(c). We find that this strain induced band modulation in the conduction band is almost an order larger than that in the valence band, which is consistent with previous calculations \cite{Feng2012}. In the following, we first focus on the energy profile of the conduction band of monolayer WS$_2$. The magnitude of this modulation reaches as high as ${\sim}200\ \mathrm{meV}$, as has been experimentally observed by recent STM study \cite{Li2022}. More importantly, the local minima of this energy profile are located at $R^X_h$ and $R^M_h$ stackings with a small energy difference of $13\ \mathrm{meV}$, which suggests that the electrons should be localized around these two stackings. However, our calculation of wavefunction in Fig. \ref{fig2}(b) shows that electrons are only localized around $R^M_h$ stacking. We note that the flat miniband above lowest conduction one is still localized around $R^M_h$ stacking, which is attributed to the spin splitting of conduction band in monolayer WS$_2$ with an energy separation of ${\sim}30\ \mathrm{meV}$ \cite{Liu2015}. Therefore additional contributions are needed to explain the localization behavior of these miniband-edge states. 

\begin{figure}[tbp]
	\centering
	\includegraphics[width = 1.02\columnwidth] {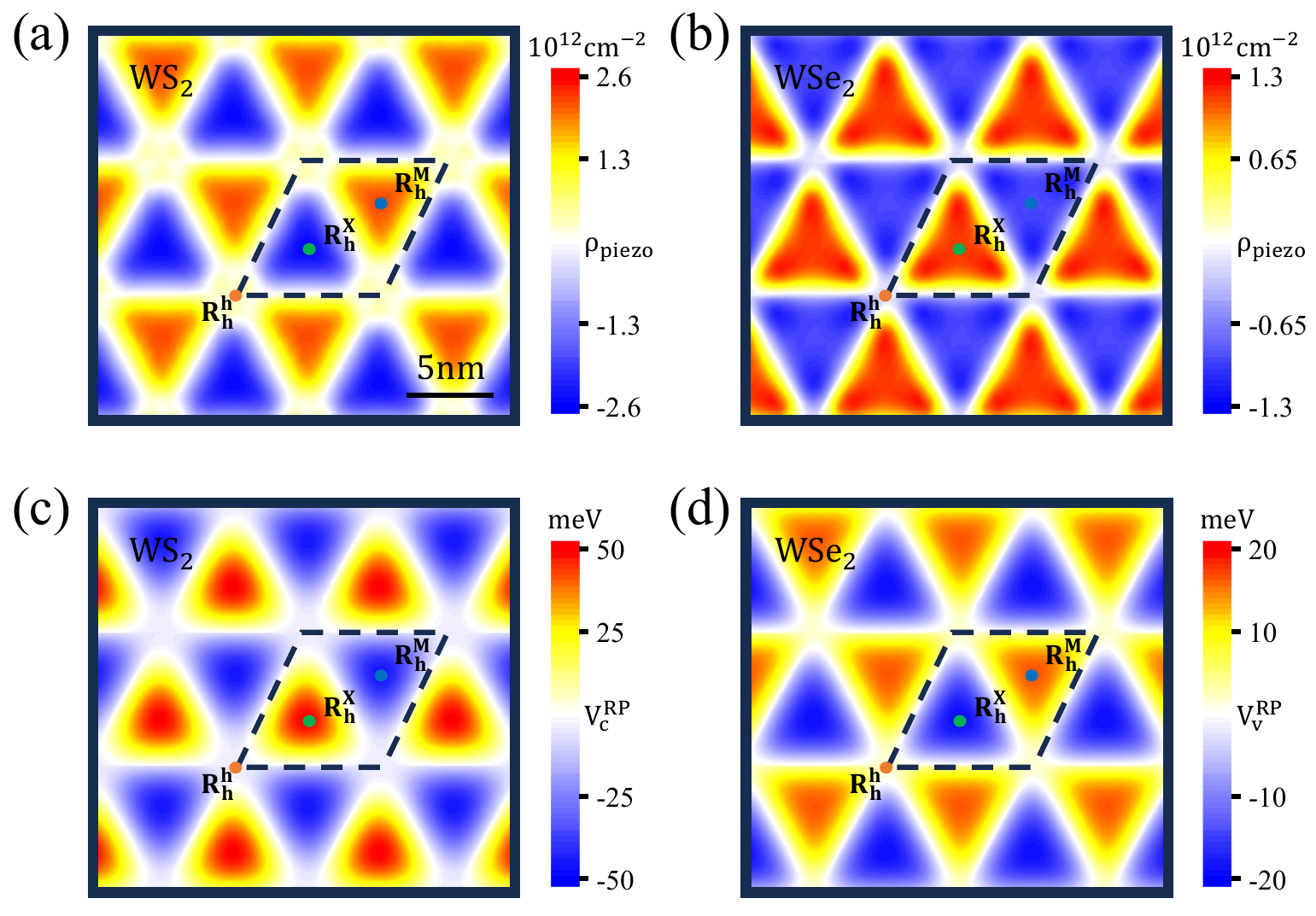}
	\caption{(a)(b) Lattice-reconstruction-induced piezocharge densities $\rho_{\text{piezo}}$ in the (a) WS$_2$ and (b) WSe$_2$ layer respectively. (c), (d) Piezopotential energy induced by piezoelectric charge shown in (a) and (b) with the screening effect considered.} \label{fig5a}
\end{figure}

Because of inversion symmetry breaking in monolayer TMDs, the inhomogeneous strain induced by lattice reconstruction can result in a piezoelectric effect \cite{Enaldiev2020,Enaldiev2D2021,Magorrian2021,Ferreira2021}. The piezocharge density is given by
\begin{equation}
	\rho_{\text{piezo}}^{(\ell)}(\mathbf{r}) = e_{11}^{(\ell)} \left[ 2\,\partial_x u_{xy}^{(\ell)}+\partial_y (u_{xx}^{(\ell)} - u_{yy}^{(\ell)}) \right],
\end{equation}
where $\ell=\{t,b\}$ is the layer index and $|e_{11}|=\{2.03 \times 10^{-10}\,\text{C/m} \ \text{for WSe$_2$,}\ \ 2.74 \times 10^{-10}\,\text{C/m} \ \text{for WS$_2$}\}$ is the piezoparameter \cite{Rostami2018}. The strain tensor is defined as the spatial gradients of the displacement field, i.e. $u_{ij}=\frac{1}{2}(\partial_iu_j+\partial_ju_i)$. The spatial distributions of $\rho_{\text{piezo}}$ in the two layers are given in Figs. \ref{fig5a}(a) and \ref{fig5a}(b). This piezocharge density further induces a screening one $\rho_{\text{ind}}^{(\ell)}=\alpha_{\text{2D}}^{(\ell)}\nabla^2_{\textbf{r}}\phi_{\text{piezo}}^{(\ell)}$, where $\phi_{\text{piezo}}^{(\ell)}$ is the electric potential produced by the piezocharges. The in-plane 2D polarizability of the two monolayer is related to the static in-plane dielectric permittivity $\epsilon_{\|}=\{15.9 \ \text{for WSe$_2$,}\ \ 14.4 \ \text{for WS$_2$}\}$ and interlayer distance $d_0=6.3\ \text{\AA}$ as, $\alpha_{\text{2D}}^{(\ell)}=d_0(\epsilon_{\|}-1)/4\pi$ \cite{Cudazzo2011}. The total potential $\phi^{(\ell)}$ is then obtained by solving the Poisson equation. Figures \ref{fig5a}(c) and \ref{fig5a}(d) show the distribution of potential energy $V_{(\ell)}^{RP}=-e\phi^{(\ell)}$ induced by this piezoelectric effect. We find that the magnitude of this potential reaches as high as ${\sim}90\ \mathrm{meV}$ in the WS$_2$ layer with the minimum located at $R^M_h$ stacking, while it is ${\sim}40\ \mathrm{meV}$ in the WSe$_2$ layer with the maximum located at $R^M_h$ stacking as well.

\begin{figure}[tbp]
	\centering
	\includegraphics[width = 1.02\columnwidth] {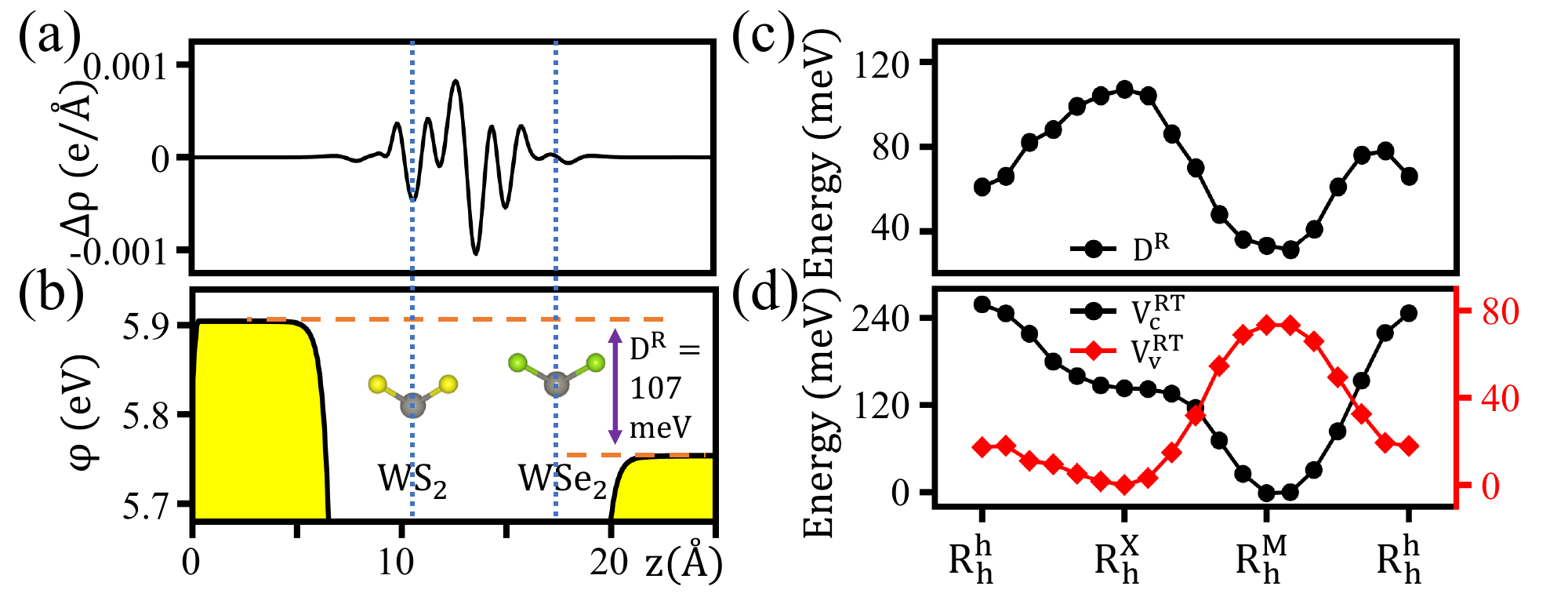}
	\caption{(a) Plane-averaged charge density difference $\Delta\rho$ for a commensurate $R^X_h$ stacking. (b) Electrostatic potential $\varphi$ obtained from the first-principles calculations including the contributions from the ionic and Hartree potentials. The charge redistribution results in a vacuum level difference $D^R$ between the two layers. (c) $D^R$ as a function of stacking orders adopted along the long-diagonal of an R-type moir\'{e} in Fig. \ref{fig1}(a). (d) The total moir\'{e} potential $V^{RT}_{c/v}$ for the conduction band (black line with dots) and valence band (red line with diamonds) contributed by local strain and interlayer charge transfer.} \label{fig5}
\end{figure}

Generally, for a vdW layered material, there exists a finite interlayer charge transfer between the constituent layers, which is responsible for the sliding ferroelectricity \cite{Li2017}. It has been shown that this charge transfer depends on the interlayer stacking, which can contribute to the moir\'{e} potentials \cite{Tong2021}. Figure \ref{fig5}(a) shows the first-principles calculated plane-averaged charge density difference, which suggests that there is a finite interlayer charge transfer across the WS$_2$/WSe$_2$ heterointerface. This charge redistribution results in a built-in electric field, which leads to a vacuum level difference $D^R$ across the two layers as indicated in Fig. \ref{fig5}(b). This potential difference agrees well with the one obtained by solving directly the Poisson equation, which suggests that the main effect of interlayer charge transfer is the introduction of a Stack effect that shifts relatively the band structures of the two layers \cite{Tong2021}. In particular, the induced electric potentials are $V^{RC}_c=D^R/2$ for the WS$_2$ layer and $V^{RC}_v=-D^R/2$ for the WSe$_2$ layer. Importantly, this interlayer charge transfer depends on the stacking order, which further results in a stacking dependence of the built-in electric field and hence the Stack-shift energy. Figure \ref{fig5}(c) shows the vacuum level difference $D^R$ as a function of different stackings along the long diagonal of the lattice-relaxed moir\'{e} pattern, from which an energy modulation on the order of $80\ \mathrm{meV}$ is observed. In particular, with a smaller atomic shell number, the S atom is easier to attract electrons from the W atom than the Se atom. As a result, for $R^X_h$ configuration, in which the S atom of WS$_2$ is vertically aligned with W atom of WSe$_2$, the charge transfer is largest. On the other hand, for $R^M_h$ configuration, the charge transfer is smallest. For $R^h_h$ configuration, the large interlayer distance also suppresses the interlayer charge transfer. Accordingly, this potential profile shows local minimum and maximum at $R^M_h$ and $R^X_h$ site respectively. Finally, taking into account the effects from local-strain-induced band modulation, piezopotential energy and interlayer charge transfer, the total energy profile of the conduction band [$V^{RT}_c\equiv V^{RS}_c+V^{RP}_c+V^{RC}_c$ as black line with dots in Fig. \ref{fig5}(d)] gives a strong trapping at $R^M_h$ site. This well explains the localization behavior of conduction miniband-edge state shown in Fig. \ref{fig2}(b).

The above discussion of moir\'{e} potential for conduction miniband-edge state also holds for the valence miniband-edge one, which is located at WSe$_2$ layer. The red line with diamonds in Fig. \ref{fig5}(d) shows the stacking dependence of the total hole potential ($V^{RT}_v\equiv V^{RS}_v+V^{RP}_v+V^{RC}_v$) contributed from the local-strain-induced band modulation, piezopotential energy and interlayer charge transfer. We note that because of the opposite effective mass, valence band states (holes) are trapped at the local maximum of the moir\'{e} potential. The calculated total potential profile shows a local maximum at $R^M_h$ site, which well explains the localization behavior of valence miniband-edge state as shown in Fig. \ref{fig2}(c). Because the overall potential profile of $V^{RT}_v$ is much smoother than that of $V^{RT}_c$, the topmost valence miniband is relatively dispersive and, accordingly, the distribution of the wavefunction is more extended than that of conduction miniband.

\begin{figure}[tbp]
	\centering
	\includegraphics[width = 1.03\columnwidth] {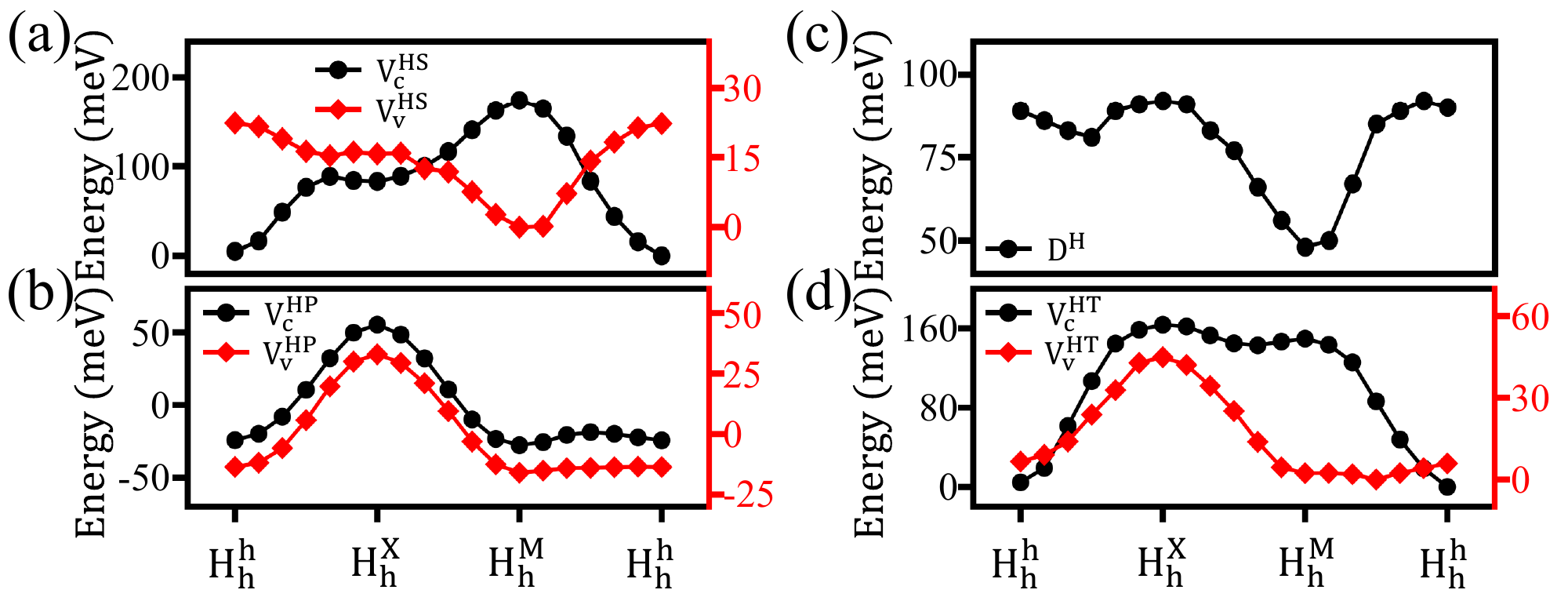}
	\caption{(a) Modulation of conduction band-edge energy of monolayer WS$_2$ ($V^{HS}_c$ as black line with dots) and valence band-edge energy of monolayer WSe$_2$ ($V^{HS}_v$ as red line with diamonds) as a function of local strain adopted along the long-diagonal of Fig. \ref{fig1}(g). (b) The piezopotential energy induced by piezoelectric charge with the screening effect considered in the WS$_2$ ($V^{HP}_c$ as black line with dots) and WSe$_2$ layer ($V^{HP}_v$ as red line with diamonds). (c) The vacuum level difference $D^H$ across the heterointerfaces with stacking orders adopted along the long-diagonal of an H-type moir\'{e} in Fig. \ref{fig1}(e). (d) The total moir\'{e} potential $V^{HT}_{c/v}$ for the conduction band (black line with dots) and valence band (red line with diamonds) contributed by local strain and interlayer charge transfer.} \label{fig6}
\end{figure}

\subsection{H-type moir\'{e}}
We now turn to studying the H-type moir\'{e}, with the focus on joint contributions from lattice-reconstruction induced local strain and interlayer charge transfer. For the conduction band, based on the local strains in WS$_2$ layer extracted from the relaxed H-type moir\'{e} pattern as shown in Fig. \ref{fig1}(g), we find a strong modulation with a magnitude of 173 $\mathrm{meV}$ [$V^{HS}_c$ as black line with dots in Fig. \ref{fig6}(a)]. This potential profile features a local minimum located at $H^h_h$ site. In addition, this lattice reconstruction induces a piezopotential energy with a magnitude of 83 $\mathrm{meV}$ [$V^{HP}_c$ as black line with dots in Fig. \ref{fig6}(b)], which features local minima located at $H^M_h$ and $H^h_h$ site. As in the case of R-type moir\'{e}, the interlayer charge transfer also induces a stacking dependent energy shift $V^{HC}_c=D^H/2$ for the WS$_2$ layer. In particular, Fig. \ref{fig6}(c) shows the first-principles calculated vacuum level difference $D^H$ across the heterobilayer, whose modulation is nearly half of that of R-type one. It is large at $H^X_h$ and $H^h_h$ sites with smaller interlayer distance and smallest at $H^M_h$ site with largest interlayer distance. The total moir\'{e} potential contributed by both local strain and interlayer charge transfer ($V^{HT}_c\equiv V^{HS}_c+V^{HP}_c+V^{HC}_c$) is given by the black line with dots in Fig. \ref{fig6}(d), in which the minimum is located at $H^h_h$ site. This is responsible for the localization of wavefunction of conduction miniband as shown in Fig. \ref{fig3}(b).

\begin{figure}[tbp]
	\centering
	\includegraphics[width = 1.02\columnwidth] {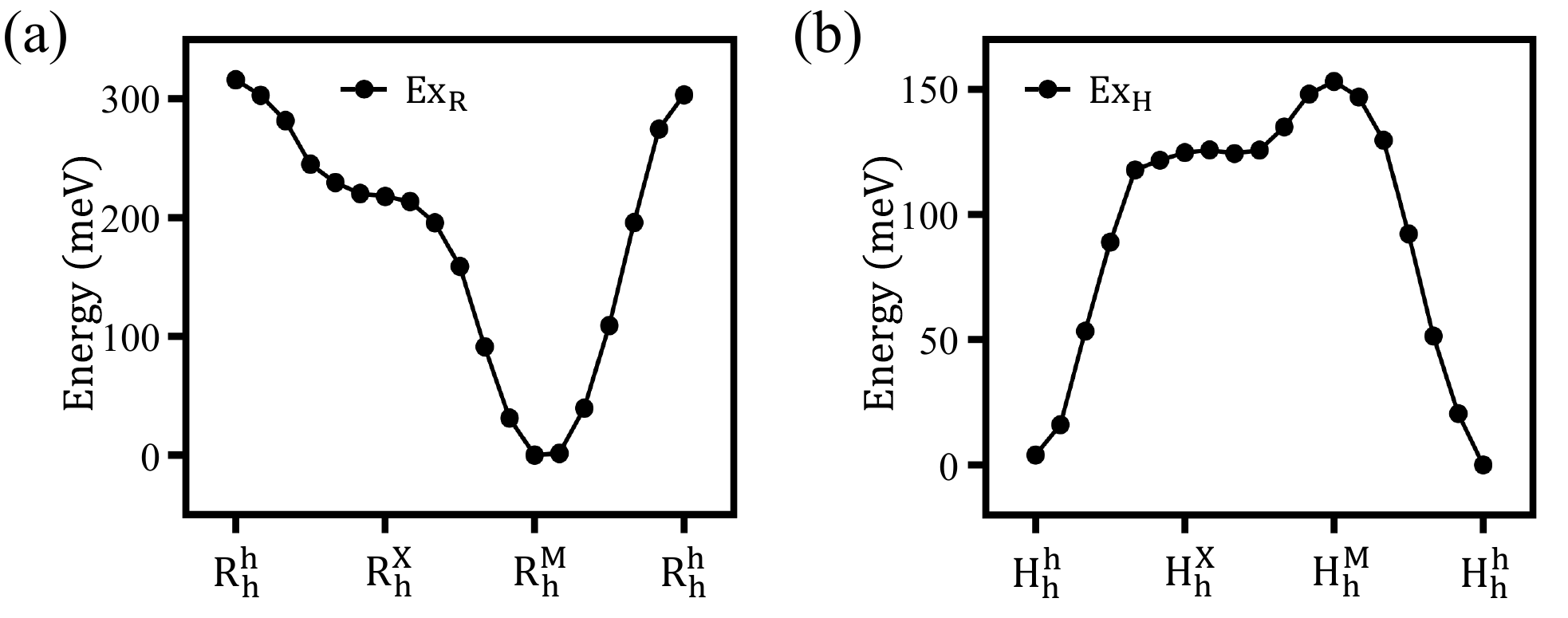}
	\caption{Excitonic moir\'{e} potential along the long-diagonal of an R-type (a) and H-type (b) moir\'{e} pattern shown in Fig. \ref{fig1}.} \label{fig7}
\end{figure}

For the valence band, the band modulation $V^{HS}_v$ induced by the local strain in the WSe$_2$ layer extracted from the relaxed H-type moir\'{e} pattern in Fig. \ref{fig1}(g) is relatively small with an amplitude of $22\ \mathrm{meV}$, as shown by the red line with diamonds in Fig. \ref{fig6}(a). We note that this potential nearly cancels with the one from interlayer charge transfer $V^{HC}_v=-D^H/2$ [cf. Fig. \ref{fig6}(c)]. The total potential ($V^{HT}_v\equiv V^{HS}_v+V^{HP}_v+V^{HC}_v$) is then dominant by the piezopotential energy [red line with diamonds in Figs. \ref{fig6}(b) and \ref{fig6}(d)], which features a  magnitude of ${\sim}50\ \mathrm{meV}$ and local maximum at $H^X_h$ site. This is consistent with the localization behavior of the valence miniband-edge state shown in Fig. \ref{fig3}(c), which is indeed localized around $H^X_h$ site.

\section{Discussions and Conclusions}\label{conclusions}
Our results show that the moir\'{e} potential is deeper for electrons than that of holes, which generates flatter minibands for the former [cf. Figs. 2(a) and 3(a)]. This agrees with typical doping-dependent photoluminescence experiments on WS$_2$/WSe$_2$ heterobilayer, where the correlation effect is more evident for electron doping than that for hole doping \cite{Huang2025,miao2021,gu2022,wang2023}. Furthermore, because the exciton energy is proportional to the difference between electron and hole energy \cite{Yu2017}, one can define the excitonic moir\'{e} potential as $Ex\equiv V_c^T-V_v^T$. Figures \ref{fig7}(a) and \ref{fig7}(b) show the excitonic moir\'{e} potentials for both R-type and H-type moir\'{e}s, which feature double local minima functioning as moir\'{e} orbital degree of freedom \cite{Jin20192,Sunx2022,Gec2023,Huang2025}. Recent polarization-resolved photoluminescence studies on H-type WS$_2$/WSe$_2$ heterobilayers have evidenced the moir\'{e}-orbital excitons localized at $H^h_h$ and $H^X_h$ stackings \cite{Huang2025}, in good agreement with our calculation shown in Fig. \ref{fig7}(b).

In summary, we have studied the origin of moir\'{e} potentials in both R-type and H-type moir\'{e} patterns in WS$_2$/WSe$_2$ heterobilayers. We show that both lattice-reconstruction induced local strain and interlayer charge transfer are needed to fully explain the localization behavior of band-edge states. In particular, the local strain results in a modulation of ${\sim}200\ \mathrm{meV}$ at conduction band, while ${\sim}20\ \mathrm{meV}$ at valence band. Furthermore, the lattice reconstruction also introduces a piezopotential energy, whose amplitude ranges from ${\sim}40$ to ${\sim}90\ \mathrm{meV}$ depending on the stacking and band-edge carrier. The interlayer charge transfer across the heterointerface also results in a modulation of band-edge state, with an amplitude of ${\sim}80\ \mathrm{meV}$ for the R-type moir\'{e} and ${\sim}40\ \mathrm{meV}$ for the H-type moir\'{e}. These contributions together well explain the localization behaviors of band-edge states for both R-type and H-type moir\'{e}s.

\begin{acknowledgments}
	This work was supported by the National Key Research and Development Program of Ministry of Science and Technology (2022YFA1204700, 2021YFA1200503), the National Natural Science Foundation of China (12374178), and the Fundamental Research Funds for the Central Universities from China.
\end{acknowledgments}

\section*{APPENDIX: THE FIRST-PRINCIPLES CALCULATIONS}\label{APPENDIXA}
The  WS$_{2}$/WSe$_{2}$ moir\'{e} superlattices contain 3903 atoms and were constructed by 26$\times$26 unit cells of WS$_{2}$ and 25$\times$25  unit cells of WSe$_{2}$, whose monolayer lattice constants are 3.17 \AA\ and 3.3 \AA, respectively. Structural relaxation was performed using the \textsc{LAMMPS} \cite{plimpton1995} package, with intralayer and interlayer interactions described by the Stillinger-Weber \cite{stillinger1985,jiang2017} potential and Kolmogorov-Crespi \cite{naik2019,kolmogorov2005} potential respectively. Atomic positions were optimized via the conjugate gradient method until the maximum atomic force was below $10^{-8}$ \,eV/\AA. Electronic structure calculations on monolayers and commensurate bilayers were based on density functional theory (DFT) \cite{kohn1965}. For the reconstructed moir\'{e} superlattice, we employed the \textsc{SIESTA} \cite{soler2002} package using a double-$\zeta$ plus polarization (DZP) basis set with a wavefunction cutoff energy of 100 \,Ry. The Brillouin zone was sampled only at the $\Gamma$ point for self-consistent charge density calculations. In the study of lattice-reconstruction induced local strain, we averaged the distances between a W atom to its six nearest-neighboring W atoms to give a local lattice constant $a$ of each monolayer. The local strain is then defined as $(a-a_0)/a_0$, where $a_0$ is the lattice constant of pristine monolayer WSe$_2$ and WS$_2$. In the study of effect from interlayer charge transfer, the stacking orders and interlayer distances were adopted from the long-diagonal of the relaxed moir\'{e} lattices. To construct the commensurate bilayer unit cells for DFT calculations, we used the average lattice constant of the relaxed monolayer WSe$_2$ and WS$_2$. All calculations were carried out with Vienna \textit{Ab initio} Simulation Package (VASP) \cite{kresse1993,kresse1996}, employing an $18\times18\times1$ \textit{k}-point mesh and an energy convergence criterion of $1\times10^{-6}$ \,eV. A vacuum spacing of 20 \,\AA\ was applied along the $z$ direction to eliminate spurious interactions between periodic images. While van der Waals corrections were essential for structural relaxation, their effects on DFT eigenvalues was negligible and thus omitted in our electronic structure calculations. Additionally, all calculations included spin-orbit coupling effects. We have checked that improving the accuracy of potential field convergence, refining the k-point mesh, increasing the plane-wave cutoff energy, changing the vdW correction scheme (for example, from DFT-D3 to optB86b-vdW), or changing the exchange-correlation function does not change noticeably the amplitude and spatial profile of the moir\'{e} potential.

\end{document}